
\input amstex
\documentstyle{amsppt}
\NoBlackBoxes
\hoffset=.75truein
\voffset=.75truein
\def\frac#1#2{{#1\over#2}}
\def\half{{\frac 12}}
\def\third{{\frac 13}}
\def\sixth{{\frac 16}}
\leftheadtext{C. G. TORRE}
\rightheadtext{A DEFORMATION THEORY OF SELF-DUAL EINSTEIN SPACES}
\topmatter
\title A Deformation Theory\\
of Self-Dual Einstein Spaces\endtitle
\author C. G. Torre\endauthor
\address Department of Physics, Syracuse University, Syracuse, New York
13244-1130\endaddress

\address Department of Physics, Utah State University, Logan, Utah
84322-4415\endaddress

\abstract The self-dual Einstein equations on a compact Riemannian 4-manifold
can be expressed as a quadratic condition on the curvature of an $SU(2)$ (spin)
connection which is a covariant generalization of the self-dual Yang-Mills
equations.  Local properties of the moduli space of self-dual Einstein
connections are described in the context of an elliptic complex which arises in
the linearization of the quadratic equations on the $SU(2)$ curvature.  In
particular, it is shown that the moduli space is discrete when the cosmological
constant is positive; when the cosmological constant is negative the moduli
space can be a manifold the dimension of which is controlled by the
Atiyah-Singer index theorem.
\endabstract

\thanks Work supported in part by NSF Grant PHY-9005790 to Syracuse
University.\endthanks
\endtopmatter

\document

\head Introduction
\endhead
The last few years have seen remarkable progress in the theory of
differentiable manifolds in 3 and 4 dimensions \cite{1}.  What is more
remarkabl
e, at
least from a physicist's point of view, is the strong link these mathematical
ideas have had with elements of field theory.  In particular, in Donaldson's
theory of 4-manifolds the moduli space of (gauge-inequivalent) solutions to the
self-dual Yang-Mills equations plays the central role.  This moduli space was
originally studied by physicists (and mathematicians too) in the context of
instanton contributions to
functional integrals in quantum gauge theory, and recently it was shown
by Witten \cite{2} how in fact a quantum field theory---``topological
Yang-Mills
theory''---provides a (necessarily somewhat heuristic) explanation for the
success of Donaldson's approach.

If Yang-Mills theory, which is firmly rooted in particle physics, should have
such a profound role to play in describing the global structure of 4-manifolds,
one is naturally led to ask: What then is the role (if any) for the earliest of
the modern geometrical theories, namely, general relativity?  The answer to
this question may lie in another remarkable recent series of results in
gravitational physics that, roughly speaking,
reveal a new way in which one can view gravitation as a
gauge theory.  What I have in mind here is the program initiated by Ashtekar
\cite {3} of
viewing the dynamics of the gravitational field in terms of the left (or right)
handed (equivalently: self-dual or anti-self-dual) spin connection.  The new
perspectives afforded by the ensuing ``connection dynamics'' point of view in
general relativity have stimulated renewed interest and fresh results in the
program of canonical quantization of the gravitational field.

As we shall see, the left-handed spin connection is also a useful variable for
studying certain aspects of classical differential geometry, in particular, the
geometry and topology of self-dual Einstein spaces.  This was realized early on
by Ashtekar, Jacobson, and Smolin \cite{4} in their study of the 3+1 form of
half-flat solutions to the Einstein equations (with vanishing cosmological
constant), and
later it was shown by Capovilla,
Dell, and Jacobson \cite{5} that the self-dual Einstein equations (with
non-vani
shing
cosmological constant)
can be expressed
purely in terms of the left-handed spin connection in a way which can be
thought of as a covariant generalization of the self-dual Yang-Mills equations.

The work that we shall present here can be viewed as a
first attempt to assess the feasibility
of applying ideas from topological field theory to the moduli space of
self-dual gravitational instantons.  Because topological quantum field theories
are nearly ``classical''---typically the semi-classical approximation is
exact---the majority of the topological field theory
formalism is dominated by the features of the
linearized theory, {\it i.e.}, the deformation theory of the
particular moduli space under consideration.  Thus in this talk we aim to
develop the deformation theory of self-dual Einstein spaces using the
self-dual spin connection as the basic variable.  We shall see that
the techniques
which were brought to bear in the corresponding Yang-Mills problem \cite{6}
can be
fruitfully applied also in the gravitational case.  At the very least, it will
become clear that the direct use of the left-handed spin connection leads to
new results in the theory of Einstein spaces at relatively little cost in the
way of extensive computations.

The geometrical setting for what follows is an $SU(2)$ principal bundle
with connection over a ``spacetime'', which will be taken to
be a compact, smooth, Riemannian, 4-dimensional spin manifold {\bf M},
and associated vector bundles equipped with covariant
derivatives.
In this framework, left-handed spinors arise as sections of the vector
bundle associated
with the defining representation of $SU(2)$.  We will primarily be concerned
with the vector
bundle constructed via the adjoint representation; the space of
smooth sections of this bundle will be denoted $S_0$ and
can be viewed as the symmetric tensor product of the bundle of left-handed
spinors with itself.  Given a soldering form, $S_0$ can be identified
with the bundle of self-dual 2-forms.  The tensor product of $S_0$
with the bundle of p-forms is denoted $S_p$.
We will use Penrose's abstract index notation \cite{13}
to describe the various geometric objects under consideration.
In particular, lowercase Latin indices will denote
tensors on {\bf M}, and
uppercase Latin indices
denote $SU(2)$ spinors, {\it i.e.},
sections of the various vector bundles. Spinor indices are lowered and raised
wi
th the
$SU(2)$-invariant symplectic form $\epsilon_{AB}$ and its inverse
$\epsilon^{AB}$.
When dealing with elements of $S_p$ it will be convenient at times to
use a matrix notation
in which the spinor indices are suppressed; in this context square brackets
$[\ ,\ ]$ will represent commutators in the Lie algebra $su(2)$.

The emphasis of this presentation will be on developing formalism and
presenting key results; no attempt will be made to be rigorous, {\it e.g.},
with
respect to functional analysis.

\head Definition of the moduli space
\endhead
The self-dual Einstein equations are
$$R_{ab}=\Lambda g_{ab}\leqno(1)$$
$$C_{abcd}=-\half\epsilon_{cd}^{\phantom{cd}mn}C_{abmn}\leqno(2)$$
where $R_{ab}$, $C_{abcd}$, and $\epsilon_{abcd}$ are the Ricci tensor, Weyl
tensor, and volume form of the metric $g_{ab}$ respectively.
$\Lambda$ is the cosmological constant. One might, more accurately, call
metrics satisfying (2) ``conformally anti-self-dual'', but for brevity we will
simply refer to them as ``self-dual''.

When $\Lambda\neq0$, eqs. (1) and (2) can be written in terms
of an $SU(2)$ (spin) connection as
follows.  We first rewrite (1) in terms of a soldering form
$\gamma_a^{AA^\prime}$,
which is an
isomorphism between vector fields and $SU(2)\times SU(2)$ spinors, and the
self-dual part of the associated spin connection $\nabla_a: S_p\to
S_{p+1}$. The
spacetime metric is obtained via
$$g_{ab}=\gamma_a^{AA^\prime}\gamma_{bAA^\prime},\leqno(3)$$
while the curvature of the left-handed spin connection is given by
$$2\nabla_{[a}\nabla_{b]}\alpha_A=F_{abA}^{\phantom{abA}B}\alpha_B,\leqno(4)$$
where $F_{ab}^{AB}=F_{ab}^{(AB)}$.
If we define the self-dual 2-forms
$$\Sigma_{ab}^{AB}:=2\gamma_{[a}^{AA^\prime}\gamma_{b]A^\prime}^B,\leqno(5)$$
which define the isomorphism between $S_0$ and the bundle of
self-dual 2-forms mentioned above,
$F_{ab}$ is related to the self-dual part of the Riemann
tensor, $R^{(+)}_{abcd}$, via
$$R^{(+)}_{abcd}=-\half F_{ab}^{AB}\Sigma_{cdAB},\leqno(6)$$
or, in an $su(2)$ matrix notation,
$$R^{(+)}_{abcd}=\half tr F_{ab}\Sigma_{cd}.\leqno(7)$$
The Einstein equations (1) are equivalent to \cite{7}
$$\nabla_{[a}\Sigma_{bc]}^{AB}=0\leqno(8)$$
and
$$\gamma_{[a}^{AA^\prime}F_{bc]A}^B +\sixth\Lambda
\gamma_{[a}^{AA^\prime}\Sigma_{bc]A}^B=0.\leqno(9)$$
Eq. (8) enforces the condition that $\nabla_a$ is the covariant derivative
coming from the left-handed (self-dual)
part of the spin
connection compatible with $\gamma_a$; given (8), (9) is equivalent to (1).

While the Einstein equations can be formulated in terms of a soldering form and
the left-handed spin connection, it is rather remarkable that the
self-dual Einstein equations can be written purely in terms of the spin
connection via
$$F_{[ab}^{(AB}F_{cd]}^{CD)}=0\leqno(10)$$
and
$$det\Phi>0,\leqno(11)$$
where $\Phi$ is a linear map from the space of symmetric rank-two
spinors to symmetric rank-two spinor densities of weight one defined by
$$\Phi^{AB}_{CD}:=\eta^{abcd}F_{ab}^{AB}F_{cdCD}.\leqno(12)$$
Here $\eta^{abcd}=\eta^{[abcd]}$ is the Levi-Civita tensor density of weight
one.

The relationship between (10),(11) and (1),(2) is as follows \cite{5,7}.
The general solution to (10) is
$$F_{ab}^{AB}=-\third\Lambda\gamma_{[a}^{AA^\prime}\gamma_{b]A^\prime}^B
=-\sixth\Lambda\Sigma_{ab}^{AB}\leqno(13)$$
where $\Lambda$, a constant with dimensions  $(length)^{-2}$, is needed for
dimensional reasons.  If we interpret $\gamma_a^{AA^\prime}$ as a soldering
form, the inequality (11) guarantees that the metric (3) is positive definite.
Now, (13) solves (9) directly, and (8) is satisfied by virtue of the Bianchi
identity:
$$\nabla_{[a}F_{bc]}=0.\leqno(14)$$
So, (10),(11) lead to a solution of the Einstein equations.  As the 2-forms
$\Sigma_{ab}$ are self-dual with respect to the metric which they define, so
too
is the $SU(2)$ field strength; it can be shown that the solutions generated in
this manned have anti-self-dual Weyl tensor.  Conversely, all anti-self-dual
Einstein spaces (with $\Lambda\neq0$) arise as solutions to (10),(11) \cite{8}.

Given a solution to (10),(11), we can generate infinitely many others by using
the
induced action of the automorphism group of the $SU(2)$ bundle being used
\footnote{I thank A. Fischer for patiently explaining to me the structure of
this
group.}.
This group, which we shall loosely call the ``gauge group'', includes the usual
local $SU(2)$ gauge group (familiar from Yang-Mills theory)
as a normal subgroup; the diffeomorphism
group of {\bf M} appears as the factor group (via bundle projection).
The space of
gauge-inequivalent solutions to (10),(11), which we shall denote $\Cal M$,
is the natural ``moduli space'' of the
problem.  Our goal in what follows is to uncover some local properties of this
moduli space by studying its tangent space $T{\Cal M}$.
\vskip 0.25in
\line{\it Remarks\hfill}

The translation of the self-dual Einstein equations into a quadratic condition
on the curvature of an $SU(2)$ connection is the result of a sequence of
observations.  Ashtekar and Renteln \cite{4} noticed that, in the context of
the
 3+1
formalism for (complex) general relativity in terms of Ashtekar's ``new
variables'', all constraints are satisfied by the ansatz
$$B^a=-{1\over3}\Lambda E^a,\leqno(15)$$
where $E^a$ is the densitized dual of the pull-back of $\Sigma_{ab}$ to a given
3-dimensional submanifold of {\bf M}, and $B^a$ is the non-Abelian magnetic
field:
the densitized dual of the pull-back of $F_{ab}$.  They also pointed out that
the evolution of such initial data sets leads to self-dual Einstein spaces.
It was Samuel \cite{7} who gave
the 4-dimensional version (13) of the ansatz and showed how it yields
solutions of the Einstein equations; shortly thereafter it was shown
that the ansatz was {\it equivalent} to the self-dual Einstein equations
\cite{8}.
Capovilla, Dell and Jacobson \cite{5} pointed out that all reference to the
soldering form could be eliminated via (10).

Because the self-dual Einstein equations can be expressed in terms of an
$SU(2)$
connection, it is easy to see that they are related to the self-dual Yang-Mills
equations.  Indeed, if we identify $-{1\over3}\Lambda E^a$ in (15) with the
Yang-Mills electric
field, then (15) is precisely the pull-back of the Yang-Mills self-duality
ansatz
to a 3-dimensional submanifold.  Furthermore, because the field
strength satisfying (13) is self-dual with respect to the metric it defines,
all solutions to (10),(11) are also solutions of the self-dual Yang-Mills
equations\footnote{For example, Samuel showed \cite{7} that the spin connection
of
De Sitter space can be identified with the single instanton configuration in
Yang-Mills theory.}
(although, of course, the converse is not true).  One can think of (10) as a
diffeomorphism covariant generalization of the self-dual Yang-Mills equations,
the latter being non-covariant because of the need for an externally prescribed
metric in their definition.

Note that there can, in principle, be topological obstructions to the
existence of solutions to (10),(11) or (1),(2).  For example, it is well-known
that {\bf M}
must have vanishing first homology if it is to admit Einstein metrics with
$\Lambda>0$. It is also easy to show that the Euler number and Hirzebruch
signature of {\bf M}
must be positive and negative semi-definite respectively.
In what follows we will indicate how some new obstructions might arise.
\vskip 0.25in
\head Deformation theory
\endhead

Let us now turn to the formal construction of the tangent space $T{\Cal M}$
to the moduli
space discussed above.  Consider a 1-parameter family of solutions
to (10),(11).  A perturbation is a tangent vector to this curve
at the point representing a given solution and is represented by an element $C$
of
$S_1$,
$$C_a^{AB}=C_a^{(AB)},\leqno(16)$$
satisfying
$$F_{[ab}^{(AB}\nabla^{\phantom{)}}_cC_{d]}^{CD)}=0,\leqno(17)$$
where $\nabla_a$ and $F_{ab}$ are built from the unperturbed solution.
The infinite-dimensional vector
space of solutions to (17) can be projected to $T{\Cal M}$ by
identifying any two perturbations $C_a$ and $C_a^\prime$ which differ by an
infinitesimal gauge transformation, {\it i.e.},
$$C_a\sim C_a^\prime$$
if
$$C_a-C_a^\prime=N^bF_{ba}+\nabla_aN\leqno(18)$$
where $N^a$ is a (complete) vector field associated with an infinitesimal
diffeomorphism of {\bf M} and $N\in S_0$
generates an infinitesimal $SU(2)$ gauge transformation.
It is straightforward to verify that any ``pure
gauge'' perturbation (18) satisfies the linearized equations (17).

The form (18) of infinitesimal automorphisms
arises as follows.  A 1-parameter family of automorphisms yields a complete
vector field on the $SU(2)$ principal bundle where the connection naturally
lives as an $su(2)$-valued 1-form; the ``infinitesimal'' action of the
automorphism group on the connection is the Lie derivative of the connection
1-form
along this vector field, which can be identified with an element of $S_1$.
Using the fixed unperturbed connection, the vector field can be split into
horizontal and vertical parts; the horizontal part yields the first term in
(18), which can be thought of as a ``gauge covariant Lie derivative", while the
vertical part of the vector field generating the automorphism leads to the
second term in (18) in the familiar way.

Crucial for the results to follow is that when $F_{ab}$ satisfies (10),(11),
and hence (13), the r.h.s of (18) can be written as
$$N^bF_{ba}+\nabla_aN=(\nabla^bf)F_{ba}+[\nabla^bL,F_{ba}]+h^b
F_{ba}+\nabla_a\hat N,
\leqno(19)$$
where $f\in C^\infty(M)$ is a real-valued function, $L,\hat N$ are elements of
$S_0$
and the background self-dual Einstein metric is used to provide the isomorphism
between vector fields and 1-forms.  Eq. (19) results from a Hodge decomposition
of $N_a$:
$\nabla_af$ is the exact part of $N_a$, $L$ comes from the co-exact part, and
$h_a$ is the harmonic part of $N_a$.  For details, see \cite{9}.  To simplify
th
e
results which follow, let us henceforth assume that {\bf M} has vanishing first
homology so that $h_a$ is in fact zero\footnote{Vanishing first homology is
guaranteed, {\it e.g.}, if {\bf M} is simply connected.}.

The tangent space to moduli space (where it is well-defined) can now be
characterized as follows.  Let $W_4$ denote the space of smooth sections
constructed as the product
of the bundle of 4-forms with the
totally symmetric trace-free
tensor product of $S_0$ with itself, {\it e.g.}, if
$\omega_{abcd}^{ABCD}\in W_4$, then
$$\omega_{abcd}^{ABCD}=\omega_{abcd}^{(ABCD)}.\leqno(20)$$
Define the following linear differential operators:
$$D_0:C^\infty(M)\oplus S_0\oplus S_0\to S_1,$$
$$D_0(f,L,N):= (\nabla^bf)F_{ba}+[\nabla^bL,F_{ba}]+\nabla_aN\leqno(21)$$
and
$$D_1:S_1\to W_4,$$
$$D_1C=F_{[ab}^{(AB}\nabla^{\phantom{)}}_cC_{d]}^{CD)}.\leqno(22)$$
Because any perturbation which is ``pure gauge'' satisfies the linearized
equations, we have
$$ D_1 D_0=0.\leqno(23)$$
The tangent space $T{\Cal M}$ to moduli space, at a given point representing
the
unperturbed self-dual Einstein space, is then simply
$$T{\Cal M}= {Ker D_1\over{Im D_0}}.\leqno(24)$$

We have thus arrived at a cohomological description of $T{\Cal M}$.
In particular, the differential complex
$$C^\infty(M)\oplus S_0\oplus S_0\ \ {\buildrel D_0
\over\longrightarrow}\ \ S_1\ \ {\buildrel D_1\over\longrightarrow}
\ \ W_4\leqno(25)$$
is elliptic, {\it i.e.}, the symbol sequence is exact, so exactly as in Hodge
th
eory we
can characterize $T{\Cal M}$ by the kernel of an elliptic differential
operator.
To do this we need inner products on the various sections which feature in the
complex.  The inner products are constructed using the unperturbed self-dual
Einstein metric and the $su(2)$ trace (equivalently: the symplectic form
$\epsilon_{AB}$), e.g., for $C, C^\prime\in S_1$ we
set
$$(C^\prime,C):=-\int_M\sqrt{g}g^{ab}trC^\prime_aC^{\phantom{\prime}}_b,
\leqno(26)$$
with obvious generalizations to sections of the other bundles.  Using this
inner
product one obtains the following adjoint operators
$$D_0^*:S_1\to C^\infty(M)\oplus S_0\oplus S_0$$
$$D_0^*C=\left(tr
F^{ab}\nabla_aC_b;[\nabla_aC_b,F^{ab}];-\nabla^aC_a\right),\le
qno(27)$$
$$D_1^*:W_4\to S_1$$
$$D_1^*\omega=F^{cd}_{CD}\nabla^b\omega_{abcd}^{ABCD},\leqno(28)$$
and ``Laplacians''
$$\Delta_0: C^\infty(M)\oplus S_0\oplus S_0\to
C^\infty(M)\oplus S_0\oplus S_0$$
$$\Delta_0= D_0^*D_0,\leqno(29)$$
$$\Delta_1:S_1\to S_1$$
$$\Delta_1= D_1^*D^{\phantom{*}}_1+D^{\phantom{*}}_0D_0^*,\leqno(30)$$
$$\Delta_2:W_4\to W_4$$
$$\Delta_2= D^{\phantom{*}}_1D_1^*,\leqno(31)$$
which are elliptic second-order partial differential operators.  In (27),(28)
we
have extended the action of $\nabla_a$ to include tensors via the connection
compatible with the background self-dual Einstein metric.

The Fredholm alternative implies the orthogonal decomposition
$$S_1=Ran D_0\oplus Ran D_1^*\oplus Ker \Delta_1,\leqno(32)$$
from which it is easy to show that
$${Ker D_1\over Im D_0}=Ker \Delta_1=Ker D_1\cap
Ker D_0^*.\leqno(33)$$
Because $\Delta_1$ is elliptic---hence Fredholm---we see that $\Cal M$ is
finite-dimensional.

Just as the alternating sum of the dimension of the kernels of Laplacians in
Hodge theory defines a topological invariant (the Euler number), the
alternating sum of the dimension of the kernels of the above Laplacians is a
topological invariant.  Let $I$ denote the ``topological index'' associated
with the complex (25) \cite{6}.  It is completely determined by the topology of
{\bf M}
and the $SU(2)$ bundle over {\bf M}.    The
Atiyah-Singer
index theorem then implies
$$I=dim Ker \Delta_0-dim Ker\Delta_1+dim Ker \Delta_2,\leqno(34)$$
so, provided $Ker \Delta_0=0=Ker \Delta_2$ (see the next section), the
dimension
of moduli space (if it exists and is a manifold) is determined by the topology
of {\bf M} via the index $I$.
\vskip 0.25in
\line{\it Remarks\hfill}

I do not yet have an explicit
expression for $I$; it is evidently a linear combination of the
Chern number $k$ of the $SU(2)$ bundle being used as well as
the Euler number $\chi$ and
modulus of the signature $|\tau|$, which are both invariants\footnote{Only the
absolute value of $\tau$ can appear because
the dimension of the kernels of the Laplacians does not depend on the
orientatio
n
of {\bf M}.}
of {\bf M}.
Because the $SU(2)$ bundle describes spinors on {\bf M}, it is possible to
relat
e the
second Chern number to $\chi$ and $\tau$ (see (6)):
$$k={1\over 2}\chi+{3\over 4}\tau,\leqno(35)$$
so $I$ can be expressed as a linear combination of $\chi$ and $|\tau|$.
Notice that, because $k$ must be an integer and $\tau$ is a multiple
of 8 for a spin manifold, $\chi$ must be an even integer.

For some purposes it is useful to rearrange the elliptic complex as follows.
Let
$$D:S_1\to C^\infty(M)\oplus S_0\oplus S_0\oplus
W_4$$
$$D:= D_0^*\oplus D_1,\leqno(36)$$
$$D^*:C^\infty(M)\oplus S_0\oplus S_0\oplus
W_4\to S_1$$
$$D^*:= D_0\oplus D_1^*.\leqno(37)$$
$D$ is an elliptic operator; the Fredholm alternative implies the
orthogonal decomposition
$$S_1=Ran D^*\oplus Ker D,\leqno(38)$$
and we have
$$T{\Cal M}=Ker D.\leqno(39)$$
The Atiyah-Singer index theorem now reads
$$I=dim Ker D^*-dim Ker D.\leqno(40)$$

A similar elliptic complex arises in the deformation theory of the moduli space
of self-dual Yang-Mills connections.  The first space in (25) is replaced by
$S_0$ (representing infinitesimal $SU(2)$ gauge
transformations, {\it i.e.}, vertical automorphisms), the second space of
sections is again $S_1$, while the third
space is the product of $S_0$ and the space of
anti-self-dual 2-forms. The topological
index in this case is again determined by $k,\chi,\tau$, which are all
independe
nt in
this case.
\vskip 0.25in
\head Vanishing theorems, linearization stability
\endhead

It turns out that the properties of the moduli space depend rather
strongly on the sign of the cosmological constant $\Lambda$.  To see this, we
shall show that, when $\Lambda>0$, $Ker \Delta_1=0=Ker \Delta_2$, while
$Ker \Delta_0=0$ when $\Lambda<0$, and explore the consequences of these
``vanishing theorems''.

Let us begin with $Ker \Delta_1$, which formally defines the tangent space to
moduli space.  We have seen that $T{\Cal M}=Ker \Delta_1=Ker D_1\cap
 Ker D_0^*$, which can be understood as expressing $T{\Cal M}$ as the
space of self-dual perturbations ($Ker D_1$) in a particular gauge ($Ker
D_0^*$).  Given that $C\in Ker D_0^*$ we have
$$\Delta_1C= D_1^*D^{\phantom{*}}_1C\leqno(41)$$
where
$$D_1C={1\over12}\epsilon_{abcd}F^{mnCD}\nabla^{\phantom{A}}_mC_n^{AB}
\leqno(42)$$
(note: in (42) we have used $D_0^*C=0$ to remove the symmetrization on spinor
indices; we have also used the self-duality of $F_{ab}$).
Explicit computation then reveals
$$\Delta_1C={\Lambda^2\over54}\nabla^b\left[\left(\delta^{[c}_a\delta^{d]}_b+
{1\over2}\epsilon_{ab}^{\phantom{ab}cd}\right)\nabla_cC_d\right],\leqno(43)$$
where we have again extended the action of $\nabla_a$ to tensors via the metric
compatible
connection of the background geometry.
Using $D_0^*C=0\Longrightarrow \nabla^aC_a=0$, it follows from (43) that
$$C\in Ker \Delta_1\Longrightarrow
\left(-\nabla^a\nabla_a+\Lambda\right)C_b=0,\leqno(44)$$
which is a remarkably simple elliptic partial differential equation for $C_b$.
It is easily seen that the linear operator $-\nabla^a\nabla_a+\Lambda$ is
positi
ve
definite when $\Lambda>0$ (relative to the inner-product introduced above) and
hence we see that $Ker \Delta_1=0$ when $\Lambda>0$.  This means that, when the
cosmological constant is positive, all perturbations of self-dual connections
are ``pure gauge'', {\it i.e.}, all instantons are isolated---the moduli space
is discrete.

A straightforward computation reveals that
$$\Delta_2\omega=0 \Longleftrightarrow
\left(-\nabla^a\nabla_a+2\Lambda\right)\omega^{ABCD}=0,\leqno(45)$$
where
$$\omega^{ABCD}:=\epsilon^{abcd}\omega_{abcd}^{ABCD}.\leqno(46)$$
Again, we see that $\Delta_2$ is a positive definite operator and $Ker
\Delta_2=0$ when $\Lambda>0$.  Finally, in a similar fashion, it can be shown
\cite{10} that $Ker \Delta_0=0$ when $\Lambda<0$.  In light of these vanishing
theorems, the Atiyah-Singer index theorem yields
$$\eqalign{&\Lambda>0: \ \
I=dim Ker \Delta_0,\cr
&\Lambda<0: \ \
I=dim Ker \Delta_2- dim Ker \Delta_1.}\leqno(47)$$

So, when $\Lambda>0$, ${\Cal M}$ is zero-dimensional and the dimension
of $Ker \Delta_0$ is
controlled by the topology of {\bf M}.  A closer look at $Ker \Delta_0$ reveals
\cite{10}
that this space can be identified with the space of Killing vectors of the
unperturbed self-dual Einstein metric.  Thus the dimension of the isometry
group of a given self-dual Einstein metric is controlled by the Euler number
and signature of {\bf M}.

When $\Lambda<0$ the dimension of moduli space is also controlled by $\chi$ and
$|\tau|$ provided that $Ker \Delta_2=0$.  When $Ker \Delta_2\neq0$ one must
confront the issue of linearization stability:  it is possible that some
solutions to (17) do not come from a 1-parameter family of solutions to
(10),(11
).
Because solutions to $\Delta_1C=0$ are to represent tangent vectors to moduli
space, spurious solutions arise when ${\Cal M}$ has singular points
where the tangent space is not well-defined.  Using the implicit function
theorem it is easy to see that a non-trivial kernel for
$\Delta_2$ represents an obstruction to the existence of a manifold structure
for ${\Cal M}$.  More precisely, the implicit function theorem implies that
$\Cal M$ exists as a manifold in the neighborhood of a self-dual instanton
provi
ded
$D$ is surjective.  From the splitting
$$C^\infty(M)\oplus S_0\oplus S_0\oplus
W_4=Ran D\oplus Ker D^*\leqno(48)$$
it is clear that $D$ is surjective
provided $Ker D^*=0$, but, when $\Lambda<0$, $Ker D^*=Ker
\Delta_2$. So, provided
$Ker \Delta_2=0$, the equations (10),(11) are linearization stable by virtue of
the
fact that ${\Cal M}$ exists (locally) as an $I$-dimensional submanifold
of the space of all $SU(2)$ connections.  Otherwise, one has to contend with
the
appearance of singularities in ${\Cal M}$ where, strictly speaking, first-order
perturbation theory fails.
\vskip 0.5in
\line{\it Remarks\hfill}

Typically, the singularities which occur in a moduli problem come from
quotienting by the action of a symmetry (gauge) group which has fixed points.
This is not the case here.  Indeed, when $\Lambda<0$, $Ker \Delta_0=Ker
D_0=0$ so there
are no fixed points.  The singularities which can occur stem from the
pathological behavior of the self-duality equations (10),(11).  To see this
note
 that
the Fredholm alternative implies the orthogonal decomposition
$$W_4=Ran D_1\oplus Ker D_1^*,\leqno(49)$$
so $D_1$ is surjective only if $Ker D_1^*=Ker \Delta_2=0.$
Hence the singularities (if any) are already present once one restricts to the
infinite-dimensional subspace (10),(11).

It is interesting to note that the sign of the topological index can represent
an obstruction to the existence of a self-dual Einstein metric (or connection)
with a given sign for the cosmological constant.  If $I<0$ then clearly (47)
cannot be satisfied when $\Lambda>0$.  Similarly, if the topology of {\bf M} is
such that $I>0$, then there can be no self-dual Einstein spaces with
$\Lambda<0$ (away from points of linearization instability).

In the deformation theory of self-dual Yang-Mills connections one can also
prove certain vanishing theorems which are
relevant to linearization stability as well as to the existence of
symmetries associated with
reducible connections.  It is also possible to have a discrete moduli space,
but this depends on the topology of the base manifold and $SU(2)$ bundle.
Singularities can
appear in the Yang-Mills version of $\Cal M$; they arise from the above
mentioned symmetries (which represent fixed points for the action of the gauge
group) and/or from the failure of the self-dual Yang-Mills
equations to define a sub-manifold.
\vskip 0.25in
\head Concluding remarks
\endhead

By studying the local properties of the space of solutions to the self-dual
Einstein equations we have seen the strong interplay between self-dual
geometry and the topology of 4-manifolds.  It is now time to assess how far we
have come toward implementing the goals expressed in the introduction, {\it
i.e.}, we should now ask: can the moduli space of self-dual Einstein
connections
tell us anything about the topology of 4-manifolds?  First of all, it is clear
that while studying the deformation theory of ${\Cal M}$ is certainly necessary
for answering this question, it is far from sufficient.  What is needed to
construct a gravitational analog of Donaldson theory is to gain control over
the
behavior of ${\Cal M}$ in the large.    It is of course going to be
a non-trivial
problem to get an analogous level of understanding of the gravitational moduli
space when
$\Lambda<0$ as one has for the moduli space of self-dual Yang-Mills
connections.  On the other hand, for manifolds admitting gravitational
instantons with $\Lambda>0$, the moduli space is discrete and one already knows
in the Yang-Mills case that a discrete moduli space leads to a new invariant
for smooth 4-manifolds.

{}From a physicist's point of view, the ``explanation'' of the success of
Donaldson theory was given by Witten via topological Yang-Mills theory.  It is
therefore encouraging to note that many features of Witten's topological
Yang-Mills theory can be reproduced in the gravitational case.  Indeed, as
shown
in \cite{11} the classical aspects of the construction of Witten's theory have
a
 natural
diffeomorphism invariant generalization to the gravitational case.  The use of
an $SU(2)$ connection to describe both gauge and gravitational instantons
leads to strong similarities between both
topological field theories and one can hope that detailed analysis
will lead to a similar degree of
success in the gravitational case as was obtained via Yang-Mills theory.
What is needed to complete the work of \cite{11}
is a better understanding of the quantum functional measure:  the existence of
a measure on connections which is consistent with the inequality (11) is bound
to be a highly non-trivial issue.  Alternatively (equivalently?), the theory
may be profitably developed using the Hamiltonian formalism and canonical
quantization.  In addition, one needs a better
understanding of the singularity structure of ${\Cal M}$, which by the way is
also a difficult issue in topological Yang-Mills theory.

We have from time to time compared the deformation theory of the space of
gravitational instantons to the corresponding deformation theory of Yang-Mills
instantons.  It is possible to give another analogy which also serves to
summarize the key results of the work presented here, namely, I would like to
argue that the moduli space of self-dual Einstein connections is a natural
4-dimensional generalization of the moduli space of Riemann surfaces.  Every
metric on a compact Riemannian 2-manifold is conformal to an Einstein metric,
where $\Lambda>0$ for genus 0, the sphere, $\Lambda=0$ for genus 1, the torus,
and $\Lambda<0$ for genus$\geq$2.    In both the gravitational and Riemann
surfa
ce
cases the moduli space is discrete when $\Lambda>0$ and the dimension of the
isometry group of the Einstein metric is controlled by the Atiyah-Singer index
theorem (which becomes the Riemann-Roch theorem in two dimensions).  When
$\Lambda<0$ the moduli space appears to be a manifold in each case whose
dimension is again controlled by the topology of the two or four dimensional
manifold via the index theorem. (The $\Lambda=0$ case, which, it seems, cannot
b
e
handled via (10), is well known---it is the K3 geometry, which can be thought
of
as the
4-dimensional generalization of the torus in Riemann surface theory.)  The
analogy between self-dual Einstein connections in 4 dimensions and Riemann
surface theory in 2 dimensions is further strengthened by the observation that
the moduli space of Riemann surfaces is identifiable with the diffeomorphism
equivalence classes of complex structures on the compact 2-manifold.
Similarly, it is easy to see that the moduli space of self-dual Einstein
connections is
closely related to the space of quaternionic K\"ahler structures on a compact
Riemannian 4-manifold \cite{12}.  More precisely, let $\sigma_{\underline{i}}\
,\ \underline{i}=1,2,3$ denote a
basis in $su(2)$, then by solving (10) and (11) one can construct three
almost complex structures,
$$J_{\underline{i}a}^{\phantom{ai}b}:=
-\left({3\sqrt{2}\over\Lambda}\right)
tr \sigma_{\underline{i}}F_a^{\phantom{a}b},\leqno(50)$$
satisfying the algebra of quaternions.

\vskip 0.25in
I would like to thank A. Ashtekar and J. Samuel for discussions.

\enddocument